\documentclass[usenatbib]{mn2e}
\usepackage{amssymb}
\usepackage{graphicx}
\usepackage{amsmath}

\title[System parameters of EC 10246-2707]{EC 10246-2707: a new eclipsing sdB + M dwarf binary\thanks{Based on observations at the SOAR telescope, a collaboration between CPNq-Brazil, NOAO, UNC, and MSU.}}

\author[B.N. Barlow et al.]{B.N. Barlow$^{1,2}$\thanks{E-mail:bbarlow@psu.edu (BNB)}, D. Kilkenny$^{3}$, H. Drechsel$^{4}$, B.H. Dunlap$^{2}$,  D. O'Donoghue$^{5,6}$, \newauthor S. Geier$^{4}$, R.G. O'Steen$^{2}$, J.C. Clemens$^{2}$, A.P. LaCluyze$^{7}$, D.E. Reichart$^{2,7}$,   \newauthor J.B. Haislip$^{7}$,   M.C. Nysewander$^{7}$, K.M. Ivarsen$^{7}$ \\   
$^{1}$Department of Astronomy \& Astrophysics, The Pennsylvania State University, University Park, PA 16801, USA\\
$^{2}$Department of Physics and Astronomy, University of North Carolina, Chapel Hill, NC 27599-3255, USA\\
$^{3}$Department of Physics, University of the Western Cape, Private Bag X17, Bellville 7535, South Africa\\
$^{4}$Dr. Karl Remeis-Observatory \& ECAP, Astronomical Institute, Friedrich-Alexander University Erlangen-Nuremberg, \\ \hspace{1.5mm}Sternwartstr. 7, D 96049 Bamberg, Germany \\
$^{5}$South African Astronomical Observatory, PO Box 9, 7935 Observatory, Cape Town, South Africa\\
$^{6}$Southern African Large Telescope Foundation, PO Box 9, 7935 Observatory, Cape Town, South Africa\\
$^{7}$Skynet Robotic Telescope Network, Department of Physics and Astronomy, University of North Carolina, Chapel Hill, NC 27599-3255, USA}\begin{document}

\maketitle

\label{firstpage}

\begin{abstract}
We announce the discovery of a new eclipsing hot subdwarf B + M dwarf binary, EC 10246-2707, and present multi-colour photometric and spectroscopic observations of this system.  Similar to other HW Vir-type binaries, the light curve shows both primary and secondary eclipses, along with a strong reflection effect from the M dwarf; no intrinsic light contribution is detected from the cool companion.  The orbital period is 0.118\,507\,993\,6 $\pm$ 0.000\,000\,000\,9 days, or about three hours.  Analysis of our time-series spectroscopy reveals a velocity semi-amplitude of $K_{\rm 1}$ = 71.6 $\pm$ 1.7 km s$^{-1}$ for the sdB and best-fitting atmospheric parameters of $T_{\rm eff}$ = 28900 $\pm$ 500 K, log \textit{g} = 5.64 $\pm$ 0.06, and log $N(He)/N(H)$ = -2.5 $\pm$ 0.2.  Although we cannot claim a unique solution from modeling the light curve, the best--fitting model has an sdB mass of 0.45 M$_{\sun}$ and a cool companion mass of 0.12 M$_{\sun}$.  These results are roughly consistent with a canonical--mass sdB and M dwarf separated by $a$ $\sim$ 0.84 $R_{\sun}$.  We find no evidence of pulsations in the light curve and limit the amplitude of rapid photometric oscillations to $<$ 0.08\%.  Using 15 years of eclipse timings, we construct an O-C diagram but find no statistically significant period changes; we rule out $|\dot{P}| > 7.2  \times 10^{-12}$.  If EC 10246-2707 evolves into a cataclysmic variable, its period should fall below the famous CV period gap.
\end{abstract}


\section{Introduction}

Few stars seem to prefer companionship as much as the hot subdwarf B (sdB) stars.  It is widely accepted that these evolved, post-main sequence stars are the progeny of red giants that were somehow stripped of their outer hydrogen envelopes, leaving behind a 0.5 M$_{\sun}$ helium-burning core surrounded by a thin layer of hydrogen \citep{heb86}.  \citet{men76} first proposed binary interactions as a probable cause for the stripping.  Since then, various models have been constructed around this idea, most of which use the angular momentum stored in orbits to spin up and eject the progenitor's envelope.  Observations tend to support this picture, with reports of the sdB binary fraction ranging from 20\% to nearly 100\% over the past few decades (see \citealt{heb09} for a review).  Today, the exact value remains uncertain, but at least a small fraction of sdBs are thought to be single stars.

The \citet{han03} models highlight five main formation scenarios leading to subdwarf B stars: the ``first'' and ``second'' Roche lobe overflow channels, the ``first'' and ``second'' common envelope channels, and the merger of two helium white dwarfs.  More recently, \citet{cla11} proposed a channel in which a helium white dwarf and a low-mass, hydrogen-burning star merge to create a hot subdwarf.  Since binary population synthesis models predict specific distributions of orbital periods, companion types, and subdwarf masses, they may be tested by observations and measurements of these parameters.  Unfortunately, the masses of sdBs are difficult to derive; fewer than 15 have been determined to date.  Most were derived from asteroseismological analyses, and, except in the case of PG 1336-018 and 2M 1938+1946, they have not been tested using independent techniques.  Other techniques leading to masses include the analysis of double-lined spectroscopic binaries, light-curve modeling of eclipsing binaries, and transit timings of eclipsing binaries.  

This work focuses on the short-period sdB binaries, which are thought to be products of common envelope evolution.  Dynamically unstable mass transfer between the hot subdwarf progenitor, a red giant, and its lower-mass main sequence companion leads to the formation of a common envelope and, eventually, the expulsion of the red giant's outer hydrogen envelope.  This process produces a hot subdwarf with a cool main sequence companion in a tight, circular orbit.  Orbital parameters have been measured for more than 100 close binaries and show periods ranging from 0.07 to 28 days with a median near 0.61 days (see Table A.1 of \citealt{gei11b}; \citealt{cop11}). Most measured periods fall below 1 day; only a handful are longer than 10 days.  We refer the reader to Figure 5 of \citet{bar12b} for an up--to--date orbital period histogram of all sdB binaries with solved orbits.

Around a dozen sdB binaries are currently known to show eclipses; most have M dwarf companions.  Although the cool companions contribute a negligible amount to the total system flux, these binaries show both primary eclipses (when the M dwarf occults the sdB) and secondary eclipses (when the sdB occults the M dwarf); in the latter case it is the reflection effect from the companion which is occulted.  It is widely believed these systems will evolve into catacslysmic variables (CVs) once the orbits decay sufficiently from angular momentum loss (see \citealt{war03} for a review).  \citet{qia08}, however, present evidence showing the orbits might evolve so quickly that mass transfer begins before the hot subdwarf has evolved into a white dwarf, leading to something other than a cataclysmic variable.  Long-term monitoring of eclipse timings will help shed light on their evolutionary rates.  Since modeling the binary light curves can lead to estimates of the component masses, each additional eclipsing sdB system discovered helps improve our understanding of the sdB mass distribution of common envelope-produced binaries.

Here, we present estimates of the system parameters for the new eclipsing sdB+dM system EC 10246-2707 (hereafter, EC 10246; $\alpha=10^h 26^m 56.59^s$, $\delta=-27^{\circ} 22' 58.73''$, J2000; B=14.2), which was discovered to be an sdB star in the Edinburgh-Cape survey \citep{ecsurvey}.  Two of us (BNB \& BHD) discovered photometric variations in EC 10246 on 21 May 2009 while carrying out a survey for new pulsating sdB stars with the 0.4-m Panchromatic Robotic Optical Monitoring and Polarimetry Telescopes (PROMPT) on Cerro Tololo in Chile \citep{rei05}.  Unbeknownst to them at the time, two of the other co-authors had already discovered eclipses at SAAO (DO'D) and had been observing the system over several seasons (DO'D and DK).  Using simultaneous spectroscopy and multi-colour photometry, we derive light-curve modeling solutions for the system and draw comparisons to other known eclipsing sdB+dM binaries.

In \S \ref{sec:spec}, we present time-series spectroscopic observations and the radial velocities and atmospheric parameters derived from them.  Photometry of the system is described in \S \ref{sec:phot}, along with an orbital ephemeris and the details of our binary light curve modeling.  \S \ref{sec:system} follows with a presentation of the best-fitting system parameters derived from this modeling.  Finally, we discuss several miscellaneous items in \S \ref{sec:discussion}, including similarities with other eclipsing sdB+dM binaries, orbital period changes, the subsequent evolution of EC 10246, and a newer method for obtaining independently-derived masses of the individual components.  We conclude with a brief summary of our work in \S \ref{sec:conclusions}.

\section{Time-resolved Spectroscopy}
\label{sec:spec}
\subsection{Observations and reductions}
We obtained 228 low-resolution spectra of EC 10246 with the Goodman Spectrograph \citep{cle04} on the 4.1-m SOuthern Astrophysical Research (SOAR) telescope on 15 December 2009.  We used a 10\arcsec\ slit with the hopes of creating a light curve from our spectroscopy.  With such a large slit, the spectral resolution is set not by the slit width but by the seeing, which ranged from 1.0-1.5\arcsec .  These conditions in combination with the 600 mm$^{-1}$ VPH grating (1.3 \AA\ per binned pixel dispersion) gave us an average spectral resolution of 5.4 \AA\ over the wavelength range 3600-6276 \AA.  The slit axis was aligned to 216.6 $\deg$ E of N, allowing us to place a second star on the slit (1.22\arcmin\ away) to monitor drifts in the wavelength solution over the run length.   We read out only a subsection of the CCD (approximately 2071$\times$623 binned pixels) to minimize the time between exposures.  Each spectrum was integrated for 40 s, resulting in a signal-to-noise (S/N) ratio of approximately 20 per resolution element.  The entire series lasted 3.1 hours, covering slightly more than one orbital cycle.  Table \ref{tab:spec_log} presents a summary of the observational setup.  HgAr comparison spectra were obtained before and after the series.  Using the same instrumental setup, we also took four 30-s spectra of the spectrophotometric standard star EG 21 for flux calibration.

\begin{table}
\caption{Time-series Spectroscopy Log}
\centering
\begin{tabular}{rl}
\hline
\hline
UT Date & 15 Dec 2009 \\
Start Time$^{a}$ & 5:15:02.8 UTC\\
Run Length & 3.1 hr\\
Grating & 600 mm$^{-1}$ \\
Slit Width & 10\arcsec \\
Dispersion & 0.65 \AA\ pixel$^{-1}$\\
Resolution & $\approx$ 5.4 \AA\\
Wavelength Range & 350-620 nm\\
Exposure Time& 40 s\\
Dead Time & 9 s\\
Cycle Time & 49 s\\
Duty Cycle & 82\%\\
No. exposures & 228\\
\hline
\multicolumn{2}{l}{$^{a}$mid-exposure time}\\
\end{tabular}
\label{tab:spec_log}
\end{table}

We used the \textit{ccdproc} routine in {\sc iraf} to bias-subtract and flat-field all spectral images and \textit{apall} to optimally extract one-dimensional spectra from the frames and subtract a fit to the sky background.  The HgAr frames we obtained were insufficient for wavelength calibration since the emission lines were too broad (from the big slit) to accurately centroid.  Instead, we used a dispersion solution from a different night created from a HgAr lamp observed through a smaller slit.  Unfortunately, this makes it impossible to measure absolute space velocities, although relative measurements can still be made.  We applied the same reduction techniques to the EG 21 spectra and used them to flux-calibrate the EC 10246 data.  The spectrum is dominated by hydrogen Balmer absorption features and also shows several helium I lines (4026 \AA, 4471 \AA, 4922 \AA, 5876 \AA).  An sdOB classification is ruled out due to the absence of the helium II line at 4686 \AA.

\subsection{Radial velocity curve}

\begin{figure}
  \centering
  \includegraphics{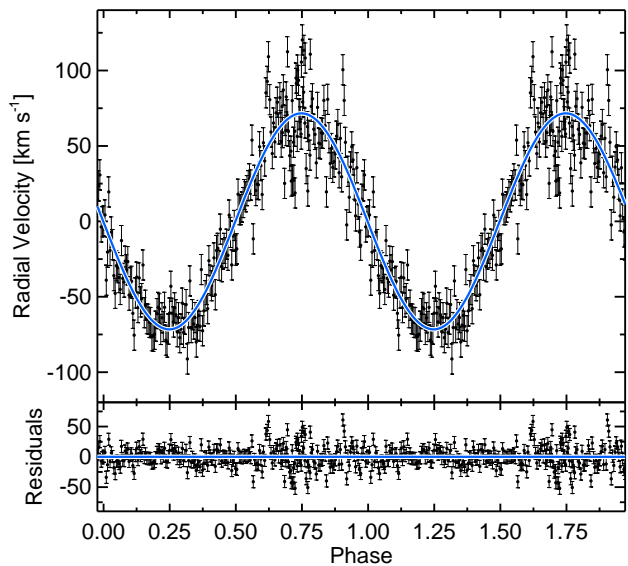}
   \caption{\textbf{\textit{Top panel:}} Radial velocity curve for EC 10246, plotted twice for better visualization.  Our full velocity curve covers 1.1 orbital cycles, but to simplify the presentation of the data, we only show one continuous cycle here.  Velocities were measured from the hydrogen Balmer profiles in the sdB spectra. The solid line denotes the best-fit sine wave to the data.  \textbf{\textit{Bottom panel:}}  Residuals after subtracting the best-fit sine wave from the data.  The mean noise level in the Fourier transform of the residual velocity curve is 2.2 km s$^{-1}$.  } 
  \label{fig:RV}
\end{figure}

Relative velocity shifts were computed from the cores of the hydrogen Balmer profiles.  We used the {\sc MPFIT} routine in IDL  \citep{mar09}, which employs the Levenberg-Marquardt method, to fit inverse Gaussians to H$\beta$-H9.  The helium I lines were too weak in the individual spectra to be used for this purpose.  Since the telescope guide star was significantly redder than EC 10246, our slit alignment was not constant over the observing run, resulting in a gradual, colour-dependent velocity shift.  By approximating the colour of the guide star, we were able to remove this wavelength-dependent shift.  Velocity drifts due to instrumental flexure were removed by tracking the absorption-line features in the companion star on the slit.  Figure \ref{fig:RV} shows the resulting RV curve for the sdB, plotted twice for better visualization.  The larger scatter near the peak of the curve resulted from diminishing weather conditions towards the end of our observing run.

We determined the semi-amplitude of the velocity variation ($K_{\rm 1}$) by fitting a sine wave to the data with {\sc MPFIT} and fixing the orbital period and phase to the values determined from the eclipse timings (see \S \ref{subsec:ephemeris}).   We derive a line-of-sight radial velocity semi-amplitude of $K_{\rm 1}$ = 71.6 $\pm$ 1.7  km s$^{-1}$ for the sdB component.  Residuals from subtracting this fit, shown in the bottom panel of Figure \ref{fig:RV}, are consistent with random noise.  We also fitted eccentric models to the data (with low eccentricity), but currently have no reason to prefer them over the circular model.  

\subsection{Atmospheric parameters}
\label{sec:spec_fit}

\begin{figure}
  \centering
  \includegraphics[scale=0.9]{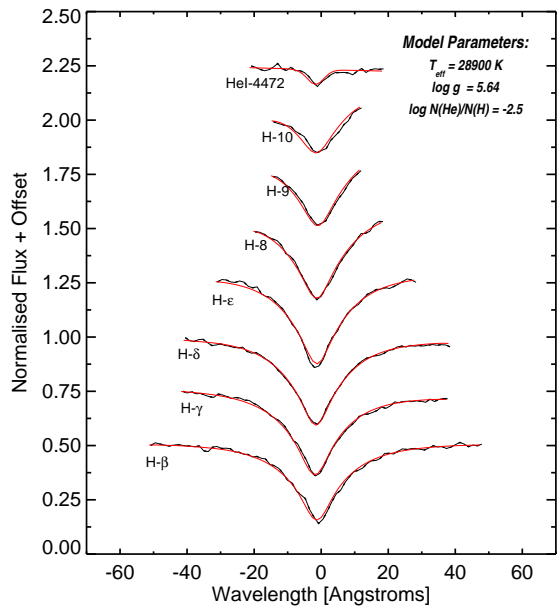}
   \caption{Normalised spectrum of EC 10246 (black dot-dashed line) with the best-fitting model (solid red line).  Absorption line profiles are offset by 0.25 from one another.} 
  \label{fig:spectrum}
\end{figure}

We fitted grids of metal-line blanketed LTE model atmospheres with solar metallicity \citep{heb00} to the spectra with the Spectrum Plotting and Analysis Suite (SPAS; \citealt{hir09}) to determine the effective temperature, surface gravity, and helium abundance of the hot subdwarf component.  As some sdB+dM binaries have spectroscopic parameters that reportedly vary over each orbital cycle (see for example, \citealt{heb04,for10}), we analysed the data according to their orbital phase.   We divided the time series into 10 phase bins, averaged the spectra in each bin, and fitted atmospheric models to the resulting spectra.  Hydrogen Balmer lines H$\beta$-H10 and the 4471 \AA\ helium I line were fitted with a $\chi^2$ minimization technique.  The best-fitting atmospheric parameters showed no change with orbital phase.  Overall, we adopt the weighted average of their values and report $T_{\rm eff}$ = 28900 $\pm$ 500 K, log \textit{g} = 5.64 $\pm$ 0.06, and log \textit{N}(He)/\textit{N}(H) = -2.5 $\pm$ 0.2.  
In Figure \ref{fig:spectrum}, we show the best-fitting solution with the average observed spectrum over one of the 10 phase bins.  The best--fitting sdB temperature cited above is used as a fixed parameter in the light-curve modeling discussed in \S \ref{sec:modeling}.

\section{PHOTOMETRY}
\label{sec:phot}
\subsection{Observations and reductions}
Simultaneously with the SOAR spectroscopy, we used PROMPT 1, 3, 4, \& 5 to obtain approximately 3.7 hrs of time-series photometry with the Johnson BVRI filters.  Although these telescopes are primarily dedicated to GRBs and science education, they are eminently useful for studies of rapidly pulsating stars and eclipsing binary systems such as EC 10246.  Cycle times ranged from 40 s (B) to 20 s (I) and duty cycles from 90\% (I) to 80\% (B).  On the following night, we took three hours of photometry through the B filter with SOAR/Goodman using 15-s exposures and an 81\% duty cycle.  Together, these five datasets represent the photometry used in our light-curve modeling.  We obtained numerous other light curves with both PROMPT and the SAAO 1.0-m telescope for the purpose of constructing the ephemeris and O-C diagram shown in \S \ref{subsec:ephemeris}.  

Reduction of the SAAO CCD frames was performed on--line, enabling the
observer to judge the quality of the observations and to select suitable
stars as local comparisons (to correct for small transparency variations).
Conventional procedures (bias subtraction, flat field correction, etc.)
were followed, with magnitude extraction being based on the DoPHOT program 
described by \citet{sch93}.

All PROMPT frames were bias-subtracted and flat-fielded using the \textit{ccdproc} procedure in {\sc iraf}.  Due to non-negligible dark current in these data, we also dark--subtracted the frames in {\sc iraf}.  We extracted photometry with the \textit{hsp\_nd}\footnote{written by Antonio Kanaan} routine by choosing apertures that maximized the signal-to-noise ratio in the light curves.  Sky annuli were drawn around the stellar apertures to keep track of and remove sky counts.  We also extracted photometry of nearby, bright comparison stars to remove sky transparency variations and normalised the target light curves with parabolas to take out atmospheric extinction effects.   The resulting differential light curves from SOAR and PROMPT used to model the binary are shown in Figures \ref{fig:SOAR_lcs} and \ref{fig:PROMPT_lcs}, respectively.

\subsection{Orbital ephemeris}
\label{subsec:ephemeris}

\begin{table}
\centering
\caption{Primary Eclipse Times of Minima}
\begin{tabular}{cccl}
\hline
\hline
Time of Minimum  &  Telescope & O-C & Notes\\
  (BJD$_{TDB}$) &  & (s)  &  \\
\hline
  2450493.46728 &            SAAO &   7.6 &   \\
  2450494.41537 &            SAAO & -15.1 &   \\
  2450494.53382 &            SAAO &   6.5 &   \\
  2450512.42856 &            SAAO &  13.2 &   \\
  2450554.38046 &            SAAO & -23.0 &   \\
  2450558.29116 &            SAAO &  -8.2 &   \\
  2450559.23922 &            SAAO & -12.2 &   \\
  2452285.54523 &            SAAO &  -2.0 &   \\
  2452350.36901 &            SAAO &   1.7 &   \\
  2452728.29108 &            SAAO &   0.9 & poor conditions  \\
  2452728.40958 &            SAAO &  -7.7 & poor conditions  \\
  2452730.30569 &            SAAO &   1.1 &   \\
  2452730.42420 &            SAAO &  -2.4 & poor conditions  \\
  2453413.38580 &            SAAO &  -5.7 & poor conditions \\
  2454974.49160 &            PROMPT &  10.7 &   \\
  2455151.77930 &            PROMPT &   2.8 &   \\
  2455162.80080 &            PROMPT &   3.9 &   \\
  2455180.69560 &            PROMPT &   0.9 &   \\
  2455180.81370 &            PROMPT &  -0.6 &   \\
  2455182.82850 &            PROMPT$^a$ &   1.8 & \\
  2455183.22850 &            SOAR &   0.3 & \\    
   2455190.76850 &             PROMPT &   5.6 &   \\
  2455212.69260 &             PROMPT &   1.4 &   \\
  2455664.68200 &             PROMPT &   3.3 &   \\
  2455664.68202 &             PROMPT &  -4.5 &   \\
  2455667.52629 &            PROMPT &  -3.5 &   \\
  2455667.52630 &            PROMPT &  -4.7 &   \\
  2455667.64476 &             PROMPT &   0.6 &   \\
  2455667.64480 &             PROMPT &   7.5 &   \\
  2455677.48108 &             PROMPT &   5.0 &   \\
  2455677.59949 &             PROMPT &  -5.7 &   \\
  2455680.56221 &             PROMPT & -10.7 &   \\
  2455692.53148 &             PROMPT & -16.6 &   \\
  2455693.47952 &             PROMPT &   8.9 &   \\
  2455693.59806 &             PROMPT &  13.1 &   \\
  2455706.51545 &             PROMPT &   9.6 &   \\
  2455715.28506 &             PROMPT &   2.0 &   \\
  2455715.28505  &           SAAO &4.2 & \\
  2455716.47003 &             PROMPT &   9.2 &   \\
  2455716.58855 &             PROMPT & -16.2 &   \\
  2455721.56594 &            PROMPT &   2.7 &   \\
  2455722.51393 &            PROMPT &   1.1 &   \\
  2455725.47675 &             PROMPT &   1.0 &   \\
  2455726.54327 &             PROMPT &   0.2 &   \\
  2455743.48999 &             PROMPT &  10.5 &   \\
  2455938.55411 &             SAAO &   4.5 &   \\
  2455940.56876 &             SAAO &   6.2 &   \\
  2456102.21372  &           SAAO & 10.7 & poor conditions\\
  2456104.22838  &          SAAO & 13.4 &\\
    \hline
    \multicolumn{4}{l}{\small $^a$ combined point from PROMPT multi--colour photometry}
\end{tabular}
\label{tab:O-C}
\end{table}

In order to compute an orbital ephemeris for EC 10246, times of primary minima were determined using two different techniques.  For the PROMPT eclipses, we fitted inverse Gaussians to the eclipse profiles using the MPFIT routine in IDL.  The SAAO minima were determined by hand via the bisected chords technique (e.g., \citealt{kil00}), which measures the mid-points of chords joining ingress and egress curves.  Table \ref{tab:O-C} presents each measured time of minimum  as a Barycentric Julian Date of the Barycentric Dynamical Time (BJD$_{TDB}$; \citealt{eas10}). From these measurements, which span more than a decade, we constructed a linear ephemeris of the form

\begin{equation}
C = T_{\rm 0} + P_{\rm 0}\times E
\end{equation}
where $T_{\rm 0}$ is a reference eclipse time, $P_{\rm 0}$ the orbital period (at $T_{\rm 0}$), and $E$ the orbital cycle number, as measured from $T_{\rm 0}$ (when $E$=0).  We assessed the quality of this model by calculating an `observed minus calculated' (O-C) diagram (see 
\citealt{ste05,kep93}).  Such a plot allows us to place limits on the orbital period change ($\dot{P}$) and to look for the presence of additional bodies in the system.

\begin{figure}
  \centering
  \includegraphics{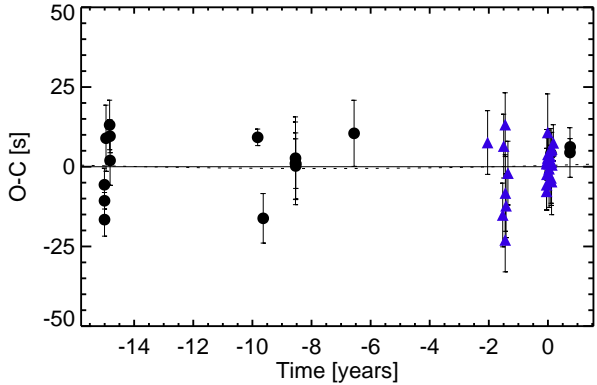}
   \caption{O-C diagram constructed from primary eclipse timings from SAAO (black circles) and PROMPT (blue triangles).   The solid and dotted lines denote the best-fitting linear and quadratic curves, respectively.  There is currently no reason to prefer a model with non-zero $\dot{P}$ over a constant-period model.  We limit secular period changes to $|\dot{P}|$ $<$ 10$^{-12}$ s s$^{-1}$.} 
  \label{fig:O-C}
\end{figure}

Figure \ref{fig:O-C} shows the full O-C diagram, from which we derive an orbital ephemeris defined by
 
 \begin{equation}
\begin{split}
BJD_{TDB} = (2\,455\,680.562\,160  \pm 0.000\, 016) \,\,\,\,\,\,\,\,\,\,\,\,\,\,\,\,\,\,\,\,\,\,\,\,\,\, \nonumber \\
+ \,\,(0.118\,507\,993\,6 \pm 0.000\,000\, 000\,9) \times E \nonumber \\
 \end{split}
 \label{eqn:ephemeris}
 \end{equation}
We fitted models with both constant and changing periods, but computation of the F--statistic shows we have no reason to prefer the best--fitting $\dot{P}$ model over a constant--period solution.  Accordingly, we claim neither a measurement of $\dot{P}$ nor the detection of tertiary companions from reflex motion.  We can, however, limit secular period changes in the system to $|\dot{P}|$ $<$ 7.2 $\times$ 10$^{-12}$ s s$^{-1}$.  Assuming a reasonable distance for EC 10246 (from the parameters derived in \S \ref{sec:modeling}) and combining this with proper motion measurements from \citet{roe10}, we expect a period increase due to proper motion on the order of 10$^{-15}$ s s$^{-1}$ \citep{shk70, paj95}, well below our detection threshold.

\subsection{Binary light curve modeling}
\label{sec:modeling}

Since only spectral features from the primary component have currently been identified, we rely on a light curve analysis to compute the mass ratio ($q$) of the binary.  To prepare the data for modeling, we normalised the light curves by the mean flux and used the calculated ephemeris to convert times to orbital phase.  We use the standard convention of defining the primary eclipse centre as the zero point in phase.  For the light-curve analysis, we employed the MOdified ROche (MORO) code, which is based on the Wilson-Devinney code \citep{wd_code}.  We refer the reader to \citet{moro} for additional details concerning MORO.  We computed separate solutions for the SOAR B curve (``Solution 1''), the PROMPT B curve (``Solution 2''), and the simultaneous PROMPT BVRI photometry (``Solution 3'').  In all cases, we fixed the binary type to `detached' in the Wilson-Devinney code (``mode 2'').  

 \begin{figure}
  \centering
  \includegraphics{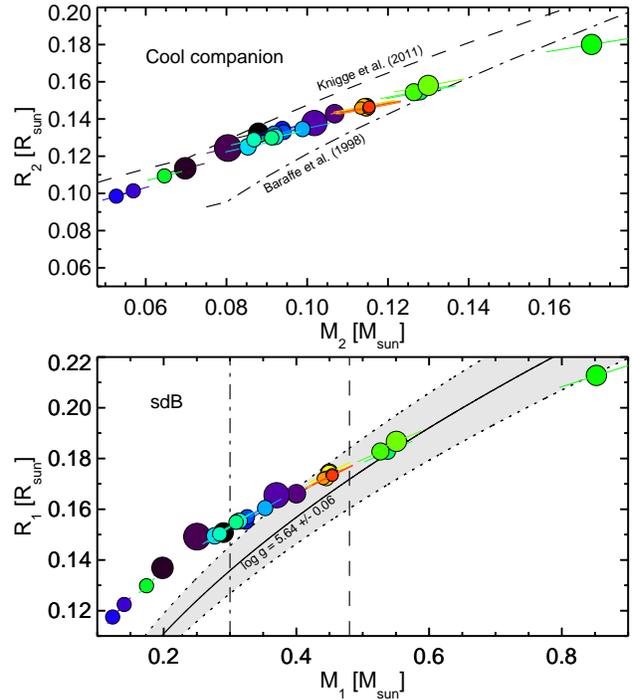}
   \caption{Mass-radius diagrams for the secondary (\textit{top panel}) and primary  (\textit{bottom panel}) stars, which illustrate the degeneracy inherent to the MORO light curve solutions.  Each pair of points represents one of the 27 light--curve modeling solutions computed from {\sc MORO}.  They are colored according to solution number to aid pair identification in the two plots; their sizes correspond to the $\chi^2$ of each model fit to the data.  The small red point denotes the best--fitting model.  In the cool companion plot,  the dot--dashed line represent the theoretical mass--radius relation for single M dwarfs from \citet{bar98}, while the dashed line represents the mass-radius relationship derived by \citet{kni11} for lower-main sequence stars in CVs. In the primary (sdB) plot, dashed and dot--dashed lines represent the `canonical' sdB mass ($\sim$ 0.48 M$_{\sun}$) and minimum mass for He-burning, respectively.  The shaded region marks our measured value of log \textit{g} with 1-$\sigma$ error bars.} 
  \label{fig:loci}
\end{figure}

\begin{figure*}
  \centering
  \includegraphics{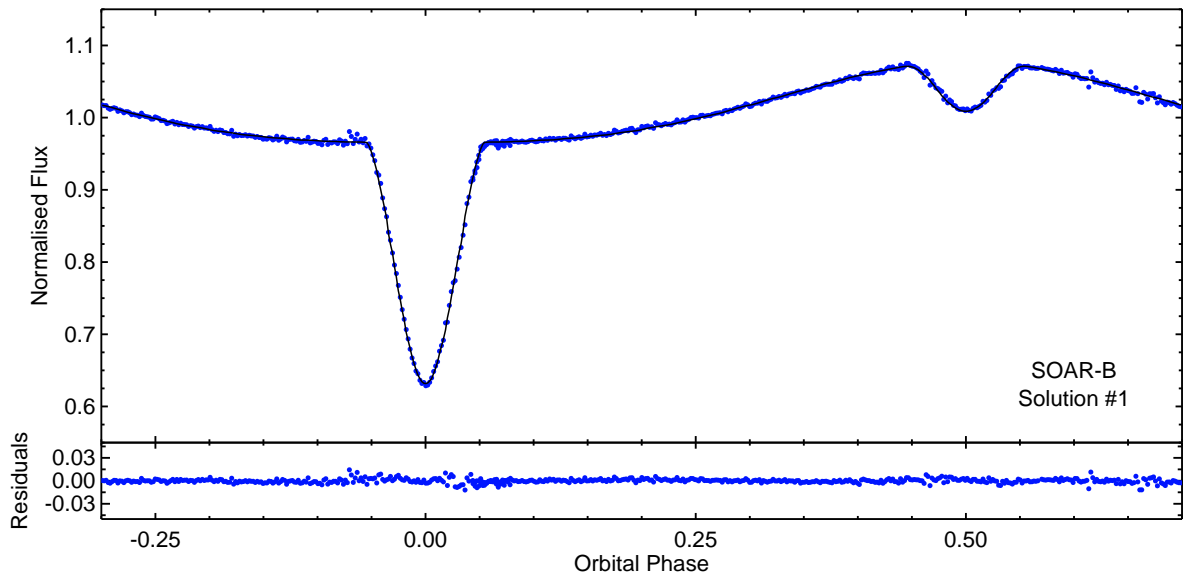}
   \caption{\textit{\textbf{Top panel:}} SOAR \textit{B} light curve with the best-fitting theoretical light curve, as determined from the MORO code.  Only one full orbital cycle of the light curve is shown.   \textit{\textbf{Bottom panel:}} Residuals after subtracting the model from the data.  } 
  \label{fig:SOAR_lcs}
\end{figure*}

\begin{figure}
  \centering
  \includegraphics{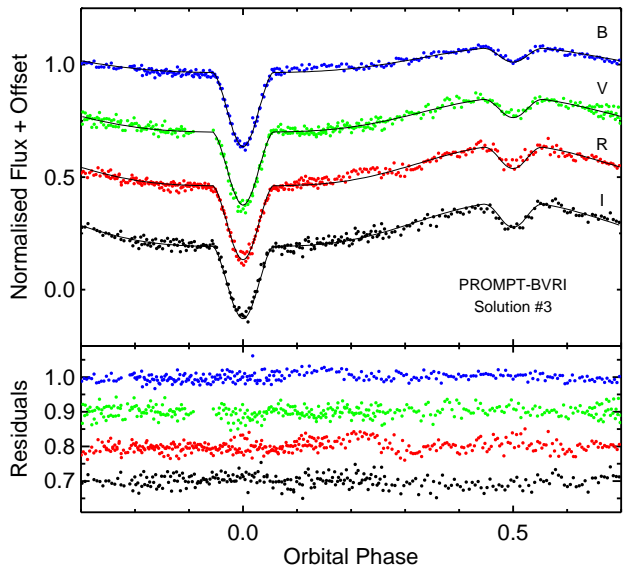}
   \caption{Simultaneous, multi-colour photometry of EC 10246-2707 obtained with PROMPT on 2009-12-15 (\textit{top panel}).  Light curves were taken through the B,V,R, and I filters with PROMPT 3, 1, 4, and 5, respectively.  Although each data set covered approximately 1.3 orbital cycles, we only show one complete cycle here.  The best-fitting binary light curves from MORO (Solution \#3) are plotted with solid black lines.  Residuals after subtraction of the models are also shown (\textit{bottom panel}). } 
  \label{fig:PROMPT_lcs}
\end{figure}

\begin{table*}
\caption{MORO Light Curve Fitting Parameters}
\centering
\begin{tabular}{lllll}
\hline
\hline
\multicolumn{5}{c}{\textbf{Fixed Parameters}}\\
\hline
Parameter & Solution \#1 & Solution \#2 & Solution \#3 & Description\\
& (SOAR-B)& (Prompt-B) & (Prompt-BVRI) &\\
\hline
$T_{\rm 1}$ [K] & 28900 & 28900 & 28900 &primary effective temperature from spectroscopy \\
$A_1$ & 1.0 & 1.0 & 1.0 &primary bolometric albedo \\
$\delta _{\rm 2}$ & 0.0 & 0.0 & 0.0 &secondary radiation pressure parameter\\
$x_{\rm 1B}$ & 0.26 & 0.26 &0.26  & primary linear limb darkening coefficient, B filter\\
$x_{\rm 1V}$ & -- & -- & 0.220 & primary linear limb darkening coefficient, V filter\\
$x_{\rm 1R}$ & -- & -- & 0.190 & primary linear limb darkening coefficient, R filter\\
$x_{\rm 1I}$ & -- &  --& 0.165 & primary linear limb darkening coefficient, I filter\\
$g_{\rm 1}$ & 1.0 & 1.0 & 1.0& primary gravity darkening exponent\\
$g_{\rm 2}$ & 0.32 & 0.32 & 0.32 & secondary gravity darkening exponent\\
$l_3$ & 0.0 & 0.0 & 0.0 &third light contribution\\
$\lambda _{\rm B}$ [nm] & 435 & 435 & 435 &isophotal filter wavelength, B filter \\
$\lambda _{\rm V}$ [nm] &  -- & -- & 555 &isophotal filter wavelength, V filter \\
$\lambda _{\rm R}$ [nm] &  -- & -- & 640 &isophotal filter wavelength, R filter \\
$\lambda _{\rm I}$ [nm] &  -- & -- & 790 &isophotal filter wavelength, I filter \\
\hline
\hline
\multicolumn{5}{c}{\textbf{Free Parameters}}\\
\hline
Parameter & Solution \#1 & Solution \#2$^a$ & Solution \#3$^a$ & Description\\
& (SOAR-B)& (Prompt-B) & (Prompt-BVRI) &\\
\hline
$q$  & 0.25 $\pm$ 0.04 & 0.253 & 0.250& mass ratio\\
$i$ [$^{\circ}$] & 79.75 $\pm$ 0.13 &  79.61 & 79.27&inclination angle \\
$T_{\rm 2}$ [K]  & 2900 $\pm$ 500 & 2861  & 2724 & secondary effective temperature\\
$A_{\rm 2}$  & 1.90 $\pm$ 0.18 & 1.84 &1.72& secondary bolometric albedo\\
$\Omega _{\rm 1}$  & 5.05 $\pm$ 0.09 & 4.9937 &4.9701& primary modified equipotential\\
$\Omega _{\rm 2}$  & 2.85 $\pm$ 0.19& 2.8457& 2.7879 &secondary modified equipotential\\
$\delta _{\rm 1}$ & 0.017 $\pm$ 0.010 &0.01971 & 0.02152 &primary radiation pressure parameter\\
$L_{\rm 1B}$ & 0.9999 $\pm$ 0.0011 & 0.99998 &  0.999992  & primary luminosity fraction$^b$, B filter\\
$L_{\rm 1V}$ & -- & -- &  0.99993  &  primary luminosity fraction, V filter\\
$L_{\rm 1R}$ & -- & -- &  0.99979  &  primary luminosity fraction, R filter\\
$L_{\rm 1I}$& -- & -- &  0.99923  &  primary luminosity fraction, I filter\\
 $L_{\rm 2B}$  & 0.00002 $\pm$ 0.00033 & 0.000016 & 0.000008 &   secondary luminosity fraction$^c$, B filter\\
 $L_{\rm 2V}$ & -- & -- &  0.00007  & secondary luminosity fraction, V filter\\
  $L_{\rm 2R}$ & -- & -- &  0.00021  & secondary luminosity fraction, R filter\\
   $L_{\rm 2I}$ & -- & -- &  0.00077  &secondary luminosity fraction, I filter\\
$x_{\rm 2B}$  & 0.601 $\pm$ 0.070 & 0.734 &0.655& secondary limb-darkening coefficient, B filter\\
$x_{\rm 2V}$  & --  & --  & 0.866& secondary limb-darkening coefficient, V filter\\
$x_{\rm 2R}$  & --  & --  & 0.998& secondary limb-darkening coefficient, R filter\\
$x_{\rm 2I}$  & --  & --  & 0.997& secondary limb-darkening coefficient, I filter\\
$r_{\rm 1}$/$a$ & 0.2058 $\pm$ 0.0013 & 0.20782&0.20833& primary radius as fraction of orbital separation  \\
$r_{\rm 2}$/$a$ & 0.1739 $\pm$ 0.0018 & 0.17407 &0.17924& secondary radius as fraction of orbital separation \\
\hline
$\sigma_{\rm fit}$ & 0.0028 & 0.0121 & 0.0160  & standard deviation about model fit\\
\hline
\multicolumn{5}{l}{$^a${\scriptsize errors are not given for Solutions 2 \& 3 since results are only derived using Solution 1.}}\\
\multicolumn{5}{l}{$^b${\scriptsize fraction of total system luminosity emitted by primary, defined as $\frac{L_{\rm 1}}{ L_{\rm 1} + L_{\rm 2}}$.}}\\
\multicolumn{5}{l}{$^c${\scriptsize fraction of total system luminosity emitted by secondary, defined as  $\frac{L_{\rm 2}}{ L_{\rm 1} + L_{\rm 2}}$.}}\\
\end{tabular}
\label{tab:fit_parameters}
\end{table*}

We use several assumptions, boundary conditions, and input from spectroscopy to constrain the parameter space searched by the models.  First, we assume the orbit is circular, a seemingly reasonable assumption given the common-envelope histories of sdB+dM binaries and their purported short circularization timescales \citep{tas92}.  We also assume the rotation of the stars is synchronised with the orbit, a process models show takes only decades\footnote{Recent observations of some sdB+dM systems (\citealt{pab12}, for example) imply the synchronization timescales might actually be longer than the EHB lifetime.} \citep{zah77}.  Since a light-curve analysis only provides fractional component luminosities, we fixed the primary temperature ($T_1$) to the spectroscopic value determined in \S \ref{sec:spec_fit} and adjusted only the secondary temperature ($T_2$).   As the radiation pressure in the M dwarf is insignificant, we set the secondary radiation pressure parameter ($\delta_2$) to zero.  The primary linear limb darkening coefficients ($x_1$) for the BVRI curve were set to the values derived by \cite{wad85}.  No significant evidence of a third light contribution ($l_3$) was found during initial investigations of the model fitting, so we kept this value locked to zero when computing final solutions.  Gravity darkening exponents of 1.0 and 0.32 were assumed for the primary \citep{von24} and secondary \citep{luc67}, respectively. Finally, the large reflection effect shows the secondary is cool while the lack of any other optical light from it implies it is small; these traits are consistent with a low--mass M dwarf companion.  Accordingly, we used the theoretical mass--radius relationship for lower--main sequence stars from \citet{bar98} to further constrain our model outputs. We accept only those MORO solutions with mass--radius combinations for the cool companion falling within conceivable boundaries obtained from \citet{bar98}.


Other adjustable parameters for the modeling included the inclination angle ($i$), secondary albedo ($A_2$), Roche surface potentials ($\Omega_1$, $\Omega_2$), mass ratio ($q$), primary radiation parameter ($\delta_1$), colour-dependent luminosities ($L_1$, $L_2$), and the secondary linear limb darkening coefficient ($x_1$).

MORO encountered no serious problems in the fitting except those connected with the usual parameter correlations. Different combinations of the inclination angle, limb-darkening coefficients, and ratio of the stellar radii, can produce essentially indistinguishable light curves of nearly the same shape and fit quality.  Thus, the code finds reasonable light curve solutions for a broad range of mass ratios ($q$=M$_2$/M$_1$) and corresponding gravitational equipotentials ($\Omega_1$, $\Omega_2$).  Figure \ref{fig:loci} illustrates this degeneracy by presenting primary and secondary mass--radius diagrams for the 27 separate solutions MORO found for the SOAR B light curve (``Solution \#1'').  

The $\chi$--squared values of the models in Figure \ref{fig:loci} are comparable, and consequently, we have no strong statistical reason to choose one model over the others without applying additional constraints.  For reference, we overplot the empirical M--R relationship for lower--main sequence stars in CVs\footnote{we note that the secondary stars in CVs might be slightly expanded compared to those in detached systems.}  from \citet{kni11}, the theoretical M--R relation for single M--dwarfs from \citet{bar98}, the `canonical' sdB mass, and the minimum mass needed for He--burning.  In the sdB panel, we also highlight combinations of masses and radii consistent with our spectroscopic measurement of the surface gravity.  Several of the models fall within this shaded region, so the addition of our log \textit{g} estimate alone is not sufficient for choosing a single solution.  In light of the degeneracy inherent to the models, we choose to report only conservative limits on the primary (sdB) and secondary (M dwarf) masses: $M_{\rm 1}$=0.35-0.85 M$_{\sun}$ and $M_{\rm 2}$=0.11-0.17 M$_{\sun}$.  

Although we cannot claim any particular solution as the definitive one, we continue our analysis and discussion using the best-fitting MORO solution (the red point in Figure \ref{fig:loci}) to illustrate a set of self--consistent parameters that might describe EC 10246.  For completeness, Table \ref{tab:fit_parameters} presents the fixed and adjustable parameters corresponding to the best-fitting MORO solutions for all data sets.  The system parameters of the best--fitting models to each of the three data sets are compatible with one another.  Throughout the rest of the manuscript, we use only Solution \#1 (the highest--quality solution) for illustration purposes.  Error bars are shown in Table \ref{tab:fit_parameters} for this solution only and represent the standard deviations of the best--fitting parameters from the family of 27 models MORO found for the SOAR light curve.

\begin{figure*}
  \centering
  \includegraphics[angle=90,scale=0.8]{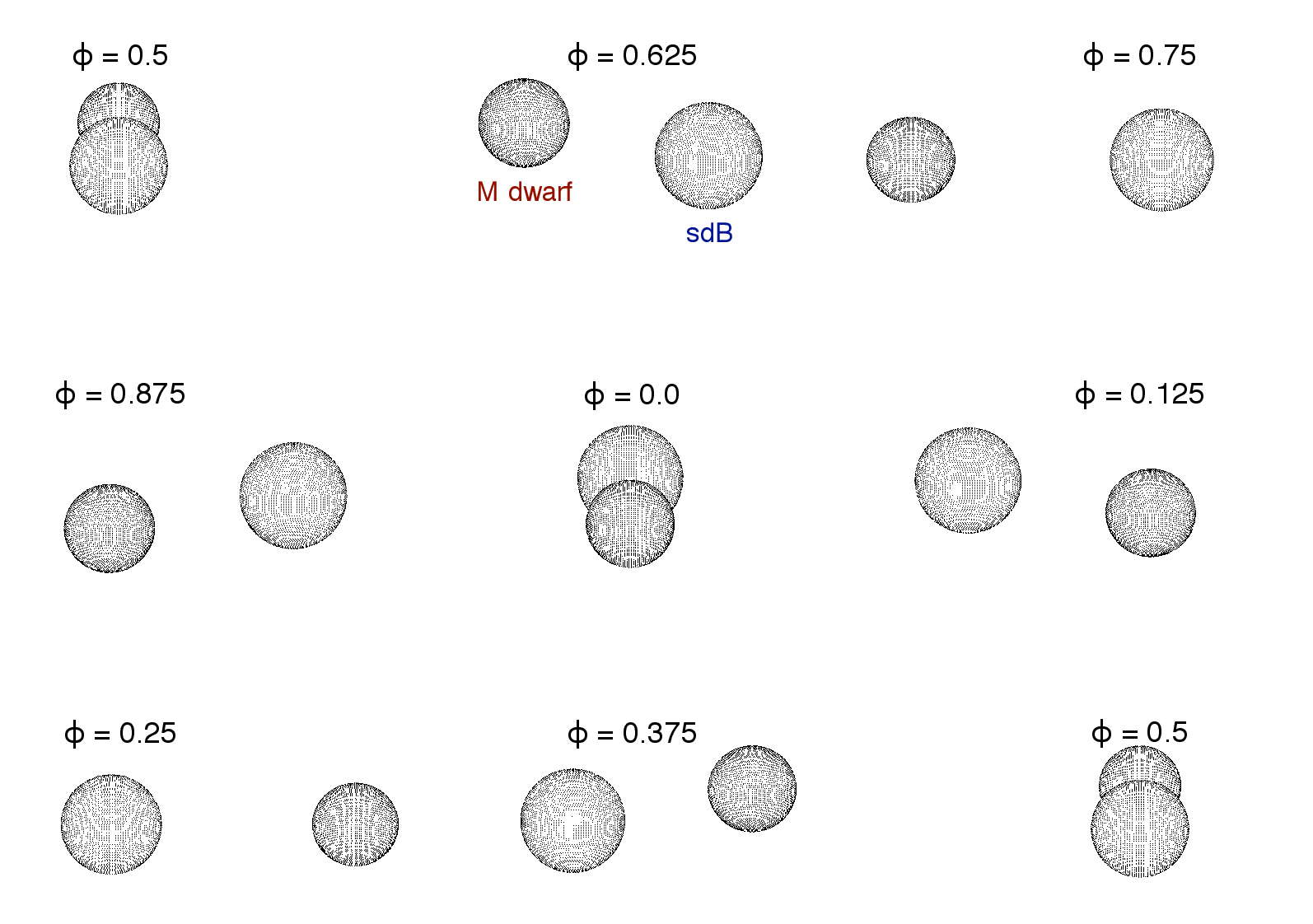}
   \caption{Snapshots from the best-fitting MORO model for EC 10246-2707 over one orbital cycle in 0.125-phase increments starting from the secondary eclipse (top left), progressing through the primary eclipse (centre image), and returning to the secondary eclipse (bottom-right; sequence progresses from left-to-right, top-to-bottom).  The system is viewed from an inclination angle of 79.8$^{\circ}$.  The separation distance and stellar radii are drawn to scale.} 
  \label{fig:snapshot}
\end{figure*}

Figure \ref{fig:SOAR_lcs} presents the best-fitting model solution to the SOAR light curve, which has a mass ratio $q$ = 0.254 and nicely fits the data.  We also show the result of the simultaneous PROMPT BVRI fit in Figure \ref{fig:PROMPT_lcs}.  Of all the output parameters, the relative radii ($r_1$, $r_2$) and inclination angle ($i$ = 79$^{\circ}$.75) are the most tightly-constrained given their strong dependence on the relatively-easy-to-measure eclipse depths and durations.  Figure \ref{fig:snapshot} shows a scaled model of the best-fit solution over one orbital cycle in 0.125-phase increments, starting at the secondary eclipse.  The large reflection effect can only be modeled using a secondary albedo $A_{\rm 2} = 1.9$, a physically unrealistic value.  Other attempts to model sdB+dM light curves have run into similar problems, which probably illustrates a mishandling of the re-radiated light in the theoretical models; to compensate for this shortcoming and still fit the light curve accurately, the secondary albedo can be adjusted.  \citet{for10}, for instance, find acceptable fits to 2M 1533+3759 only when the secondary albedo is adjusted to $A_{\rm 2} = 2.0$. The frequency redistribution in the M dwarf photosphere, which leads to emission primarily in the B filter bandpass, might produce this effect.  We discuss this further in \S \ref{sec:discussion}.

Finally, after subtracting Solution \# 1 from the SOAR light curve, we computed the Fourier transform of the residuals to look for rapid pulsations in the sdB star.  We find no significant signals in the Lomb-Scargle periodogram and limit the amplitudes of rapid, coherent photometric oscillations to $<$ 0.08\%.

\section{SYSTEM PARAMETERS}
\label{sec:system}

The mass function for EC 10246, defined by,
\begin{equation}
f  = \frac{K_{\rm 1}^3 P}{2 \pi G} = \frac{M_1 q^3 \sin^3 i}{(1+q)^2},
 \end{equation}
has a value of 0.0045 $\pm$ 0.0003 M$_{\sun}$, as calculated from the period and sdB velocity.  After inserting into the above equation the mass ratio and inclination angle from the best--fitting MORO model,  we find $M_{\rm 1}$= 0.45 $\pm$ 0.17 $M_{\sun}$ and $M_2$ = 0.12 $\pm$ 0.05 $M_{\sun}$.  The orbital separation, then, must be $a$ = 0.84 $\pm$ 0.10 $R_{\sun}$, according to Kepler's Third Law.  With this value, we can convert the radii from the MORO output, which were given in fractional units of $a$, to physical units.  We find $R_{\rm 1}$ = 0.17 $\pm$ 0.02 $R_{\sun}$ and $R_{\rm 2}$ = 0.146 $\pm$ 0.018 $R_{\sun}$, consistent with an sdB and M dwarf.  Under the assumption of synchronous rotation, we calculate a rotational velocity of $V_{\rm rot}$  = 74 $\pm$ 9 km s$^{-1}$ for the sdB component using the above orbital parameters.  Such a rotation would give rise to a 1-\AA\ broadening of the subdwarf spectral features that is four times smaller than the spectral resolution of our SOAR/Goodman spectra and, thus, difficult to measure using the current data set. Table \ref{tab:system_properties} summarizes the derived system properties for EC 10246.

We stress that $T_{\rm 2}$ is poorly constrained since the secondary contributes a negligible amount of the system light in the optical (aside from the reflection effect). Nonetheless, $T_{\rm 2}$, $M_{\rm 2}$, and $R_{\rm 2}$  are roughly consistent with the theoretical models of single lower--main sequence stars derived by \citet{bar98}, in addition to the empirical spectral type--radius relation determined by \citet{reb07} for M--dwarfs in post common envelope binaries.  Owing to the relatively large uncertainties in $M_2$, $R_2$, and $T_2$, it is difficult to determine the precise spectral type of the cool companion in EC 10246-2707; both theoretical and empirical M--R relations point to a companion near or slightly later than M5-6V.  The secondary's mass and radius are also consistent with the empirical M--R relation for CV secondaries from \citet{kni11}; we note, however, that the M dwarfs in such systems might be slightly expanded compared to detached systems such as EC 10246-2707.  Lastly, we draw attention to the close coincidence of our secondary mass and radius with those of Proxima Centauri (GJ 551) measured by \citet{dem09} using the VLTI interferometer and VINCI; they classify the star as an M5.5 dwarf.   


\begin{table}
\caption{System Properties from the Best--Fitting Solution}
\centering
\begin{tabular}{llll}
\hline
\hline
\multicolumn{4}{c}{\textbf{System Properties}}\\
\hline
$P$ & 0.118\,507\,993\,6 & $\pm$ 0.000\,000\,000\,9& days\\
$T_0$ & 2\,455\,680.562\,160 & $\pm$ 0.000\,016 & BJD$_{TDB}$ \\
$q$ & 0.25 & $\pm$ 0.05& \\
$a$ & 0.84 & $\pm$ 0.10 & R$_{\sun}$\\
$i$ & 79.75 & $\pm$ 0.13 & $\deg$ \\
$d$ & 830$^a$ & $\pm$ 50 & pc \\
\hline
\hline
\multicolumn{4}{c}{\textbf{Primary Properties}}\\
\multicolumn{4}{c}{\textit{(subdwarf B star)}}\\
\hline
$M_1$ & 0.45 & $\pm$ 0.17 & M$_{\sun}$\\
$R_1$ & 0.17 & $\pm$ 0.02 & R$_{\sun}$\\
$T_{\rm eff}$ & 28900$^{b}$ & $\pm$ 400 & K\\
log $g$ & 5.64$^{b}$ & $\pm$ 0.02 & cm $s^{-2}$\\
log{\rm $\frac{N(He)}{N(H)}$ } & -2.46$^{b}$ & $\pm$ 0.15 &\\
$K_1$& 71.6$^{b}$ & $\pm$ 1.7 & km s$^{-1}$\\
$V_{\rm rot}$ & 74$^c$ & $\pm$ 9 & km s$^{-1}$\\
\hline
\hline
\multicolumn{4}{c}{\textbf{Secondary Properties}}\\
\multicolumn{4}{c}{\textit{(M4-M5 dwarf)}}\\
\hline
$M_2$ & 0.12 & $\pm$ 0.05 & M$_{\sun}$\\
$R_{\rm 2}$ & 0.146 & $\pm$ 0.018 &R$_{\sun}$\\
$T_{\rm eff}$ & 2900 & $\pm$ 500 & K\\
log g & 5.2$^d$ & $\pm$ 0.3  & cm $s^{-2}$\\
$K_{\rm 2}$& 286$^e$ & $\pm$ 5 & km s$^{-1}$\\
$V_{\rm rot}$ & 63$^c$ & $\pm$ 8 & km s$^{-1}$\\
$A_2$ & 1.90 & $\pm$ 0.18 & \\
\hline
\multicolumn{4}{l}{$^{a}$ estimated from $T_{\rm 1}$, $R_{\rm 1}$, and extinction \citep{sch98}}\\
\multicolumn{4}{l}{$^{b}$ measured from spectroscopy (not from light curve modeling)}\\
\multicolumn{4}{l}{$^{c}$ assumes rotation synchronised with orbit}\\
\multicolumn{4}{l}{$^{d}$ calculated from $M_2$, $R_2$}\\
\multicolumn{4}{l}{$^{e}$ calculated from $q$ and $K_{\rm 1}$}\\
\end{tabular}
\label{tab:system_properties}
\end{table}

\section{DISCUSSION}
\label{sec:discussion}

In Table \ref{tab:binary_systems}, we list the parameters of all known eclipsing sdB+dM systems, ordered by increasing orbital period.  Our preliminary light curve modeling solution implies EC 10246 is a typical sdB+dM system.  Two-thirds of the M dwarfs in eclipsing sdB+dM systems have reported spectral types of M4-M5, according to their light--curve modeling solutions.  Only NSVS 14256825 (M2.5-M3.5; not well-studied), J08205+0008 ($>$M10), and HS 2231+2441 (brown dwarf; not well-studied) differ in this respect.  Whether all eclipsing sdB+dM binaries contain $\sim$ M4-M5 dwarfs remains to be seen.  It is interesting to note, however, that these spectral types correspond to the boundary at which M dwarfs become fully--convective.

Although our study is not the first to find a secondary albedo $A_{\rm 2} > 1.0$ \citep{for10}, the value we find from the best--fitting MORO solution is still disconcerting.  We can leverage HW Vir's similarity to EC 10246 to perform a consistency check by comparing their reflection effect amplitudes, similar to what \citet{for10} did for 2M 1533+3759.  We estimate the reflection effect amplitude using a simplified model in which the orbital plane is viewed exactly edge-on, both stars are perfect spheres, the secondary emits no light of its own, and the primary uniformly illuminates exactly half of the secondary's surface.  In this model, the reflection amplitude (given as a fraction of the primary flux over some wavelength range) scales with $A_{\rm 2}$$R_{\rm 2}^2$/$a^2$. 
Under the assumption that the reflection effect physics is identical for two binary systems ($j$, $k$), the ratio of their secondary albedos will be given by the expression

\begin{equation}
\frac{A_{\rm 2}^j}{A_{\rm 2}^k} =\left(\frac{R_{\rm 2}^k}{R_{\rm 2}^j}\right)^2\left(\frac{a^j}{a^k}\right)^2.
\end{equation}
From our estimates of $R_2$, and $a$, and Lee et al.'s (2009) values for HW Vir, we expect a reflection amplitude 70\% that of HW Vir in the B-filter.  The observed amplitude is 93\% of its value, implying the reflection effect actually \textit{is} stronger in EC 10246 and requires a larger albedo to model properly.  Of course, the value of $R_2$ used to demonstrate this inconsistency is somewhat dependent upon $A_{\rm 2}$, so this line of reasoning should be taken likely.  A scaling comparison to 2M 1938+4603 \citep{ost10} gives a similar result.

Eclipsing sdB+dM binaries are expected to merge eventually due to orbital angular momentum loss, primarily from the emission of gravitational radiation and magnetic braking from the M dwarf.  Once their orbits shrink sufficiently so that the secondary fills its Roche lobe, mass transfer will begin.  If the cool companion in EC 10246 truly is of spectral type M5V or later, then it is fully convective and magnetic braking should be relatively inefficient \citep{sch10}.  Consequently, gravitational radiation will dominate the rate of angular momentum loss, and the system should initiate mass transfer long after the sdB has evolved into a white dwarf and become a cataclysmic variable \citep{sch03}.  We can estimate the orbital period at this transition point ($P_{\rm cv}$) using Kepler's law.  The ratio of the current period to $P_{\rm cv}$ equals the ratio of $R_{\rm 2L}^{1.5}$ to $R_{\rm 2}^{1.5}$, where $R_{\rm 2L}$ is the current Roche lobe radius of the secondary \citep{egg83}.  We calculate a period of $P_{\rm cv}$ $\simeq$ 1.48 hrs, which falls above the minimum for CVs ($\sim$ 80 min) but below the lower-edge of the famous period gap ($\sim$ 2 hrs).

Precise eclipse measurements over a long timespan can constrain the orbital decay rates of sdB+dM binaries.  Estimates of $\dot{P}$ have been reported for many systems, including NSVS 14256825 (+1.09 $\times$ 10$^{-10}$ s s$^{-1}$, \citealt{kil12}\footnote{\citet{beu12} claim a third body in a highly elliptical orbit is responsible for the observed O-C changes}), HW Vir (-2.3 $\times$ 10$^{-11}$ s s$^{-1}$, \citealt{lee09}), and NY Vir (-1.1 $\times$ 10$^{-10}$ s s$^{-1}$,  \citealt{kil11, cam12}). For \textit{all} of these systems, however, it is probably too early to know whether the parabolic terms in their O--C diagrams are instead long-term cyclic variations.  Even with 15 years of eclipse timings for EC 10246, we are not able to claim a statistically significant $\dot{P}$ at this time.  Given the substantial amount of scatter in our O--C diagram, additional high--precision timings (e.g., \citealt{par10}) are needed to rule out small--amplitude changes, such as those claimed for NSVS 14256825 \citep{kil12,beu12}.  We note that AA Dor also shows no detectable period change after more than 35 years of monitoring \citep{kil11}.  If the physics governing orbital shrinkage is the same in all HW Vir--type systems, then the apparent orbital stability of AA Dor and EC 10246 suggests that the parabolic trends claimed in sdB+dM O--C diagrams might instead be combinations of one or more long--term cyclic changes.    

Finally, we draw attention to the opportunity eclipsing sdB+dM systems provide to measure stellar masses in a relatively model-independent way.  Due to the finite speed of light, it is possible to calculate both masses in an eclipsing binary (if M$_{\rm 1}$ $\neq$ M$_{\rm 2}$) by measuring the time delay of the secondary eclipse with respect to one half-period after the primary, a delay resulting from the extra light travel time (see \citealt{kap10}).  For a circular orbit, the delay is given by

\begin{equation}
\Delta t_{\rm LTT} = \frac{PK_{\rm 1}}{\pi c}\left(\frac{1}{q}-1\right)
\end{equation}
and increases with decreasing mass ratio.  This technique works for eccentric systems, too, but since non-zero eccentricity also affects the relative spacing between eclipses, the orbital geometry must be determined first with high precision.  The mass ratio can be computed independently of the inclination angle, while adding an estimate of this value (which is well constrained in eclipsing systems) allows one to compute the individual masses. \citet{bar12} and \citet{blo12} recently applied this method to 2M 1938+4603 (sdB+dM) and KOI-74 (A dwarf+white dwarf), respectively, using \textit{Kepler} eclipse timings to derive the component masses.  With their relatively small mass ratios and assumed circular orbits, eclipsing sdB+dM systems are good candidates for this technique.  The last column of Table \ref{tab:binary_systems} shows the predicted time delays for the known eclipsing sdB+dM systems.  Most fall near 2 s and are observable given an adequate number of precise eclipse timings (as long as the eccentricity is well-constrained).  Measurements of masses derived in this way can help test the outputs of binary light curve-modeling solutions and, in systems with pulsating sdB components (like PG 1338-018 and 2M 1938+4603), asteroseismology.


\section{SUMMARY}
\label{sec:conclusions}
We have presented photometric and spectroscopic observations of EC 10246-2707, a previously--unpublished eclipsing sdB+dM binary.  We find an orbital period of 2.84 hr and an sdB orbital velocity of 71.6 km s$^{-1}$, which together give a mass function equal to 0.0045 M$_{\sun}$.  Parameters derived from modeling the light--curve with MORO imply the hot subdwarf has a mass near the canonical value, while the cool companion is likely an M dwarf with spectral type near M5V.  The best--fitting model gives sdB and M dwarf masses of 0.45 M$_{\sun}$ and 0.12 M$_{\sun}$, respectively.  Due to numerous assumptions made during the modeling and the degeneracy associated with these light--curve solutions, however, we cannot claim a unique solution at this time. 


Like other HW Vir-type systems, the orbit of EC 10246 will decay over time, and eventually the system will become a cataclysmic variable with an orbital period below the CV period gap.  Long-term monitoring of the eclipse timings might, in due course, provide constraints on the rate of period change, as long as the effects of the system's proper motion can be identified and removed if necessary.  Timing measurements also have the potential to uncover the presence of tertiary members in the system down to planetary-size, as reported for PG 1336-018 \citep{qia11}, HW Vir \citep{lee09}, NSVS 14256825 \citep{beu12} and HS 0705+6700 \citep{beu12, cam12}.

\begin{table*}
\caption{System Parameters of Known sdB+dM Eclipsing Binaries, with sdB Properties} 
\begin{tabular}{lllllllcccl}
\hline
\hline
System Name & $P$ & $K_{\rm 1}$ & $a$ & $i$ & $M_{\rm 1}$ & $M_{\rm 2}$ & $T_{\rm eff}$ & log $g$ & $\Delta t_{\rm LTT}$ & Reference\\
(sorted by P) & [hr] & [km s$^{-1}$] & [R$_{\sun}$] & [deg] & [M$_{\sun}$] & [M$_{\sun}$] & [K] & [cm s$^{-2}$] & [s] & \\
\hline
J162256+473051$^{a}$ & 1.8 &  -- & -- & -- & -- &--&  --&  --&  --& \citet{sch11}\\
HS 0705+6700 & 2.29 & 85.8 & 0.81 & 84.4 &  0.48 & 0.13 & 28800 & 5.40 & 2.0 & \citet{dre01} \\
J08205+0008$^{\rm b}$ &  2.30  & 47.4  & 0.6/1.1 & 85.90& 0.25/0.47 & 0.045/0.068& 26700& 5.48& 2.5& \citet{gei11}\\
PG 1336-018 & 2.42 &78.6 &0.723 &80.67& 0.389 &0.110& 31300& 5.74& 1.8 & \citet{vuc07}\\
NSVS 14256825 & 2.65 & 73.4 & 0.74 & 82.5 & 0.35 & 0.097 & 42300 & 5.50 & 3.4 & \citet{alm12}\\ 
HS 2231+2441 & 2.65 &49.1 &1.18 &79.1 &0.265$^d$ &0.05$^d$ &28370 &5.39 & 2.9& \citet{ost08}\\
HW Vir & 2.80 & 82.3 &0.8594 &80.98 &0.485& 0.142& 28488& 5.60 & 2.1& \citet{lee09}\\
\textbf{EC 10246-2707$^{\rm e}$} & \textbf{2.86} & \textbf{71.6} & \textbf{0.84} & $\sim$\textbf{80}& $\sim$\textbf{0.45}& $\sim$\textbf{0.12}& \textbf{28900}&  \textbf{5.64}& \textbf{2.3}& \textbf{\textit{this manuscript}}\\
BUL-SC16 335 & 3.00 & -- & -- & -- & -- &--&  --&  --&  --& \citet{pol07}\\
2M 1938+4603 & 3.02 &65.7 &0.89 &69.45 &0.48 &0.12& 29564 &5.43& 2.3& \citet{ost10}\\
ASAS 10232-3737.0$^{\rm c}$ & 3.34 & 81.0 & 0.963 & 65.86 & 0.461 & 0.157  & 28400& 5.56 & 4.1 & \citet{sch11}\\
2M 1533+3759 & 3.88 &71.1 &0.98 & 86.6 &0.376 &0.113& 29230& 5.58& 2.5 & \citet{for10}\\
AA Dor$^{a}$ & 6.26 & 39.2 & 1.28 & 89.21 & 0.47 & 0.079 & 42000 & 5.46 &7.2  & \citet{kle11} \\

\hline
\multicolumn{11}{l}{$^{\rm a}$ unclear whether M dwarf or brown dwarf companion.}\\
\multicolumn{11}{l}{$^{\rm b}$ likely a brown dwarf, not an M dwarf.}\\
\multicolumn{11}{l}{$^{\rm c}$ marginal eclipse, large solution errors.}\\

\multicolumn{11}{l}{$^{\rm d}$ \citet{for10} argue that the log \textit{g} cited by \citet{ost08} is most likely underestimated, in which case both the sdB and} \\
\multicolumn{11}{l}{  companion are more massive than reported.}\\
\multicolumn{11}{l}{$^{\rm e}$ best--fitting solution reported; actual parameters may vary significantly.}
\end{tabular}
\label{tab:binary_systems}
\end{table*}

\section*{Acknowledgements}
We acknowledge the support of the National Science Foundation, under awards AST-0707381 (B.N.B., B.H.D., J.C.C.) and AST-0908642 (B.N.B.).  D.K. is partially funded by the South African National Research Foundation and the University of the Western Cape.  S.G. is supported by the Deutsche Forschungsgemeinschaft through grant HE1356/49-1.  R.O. received support from the NC Space Grant Consortium.  We are grateful to the SAAO for its continuing allocation of telescope time.   We recognize the outstanding observational support provided by the SOAR operators Alberto Pasten, Patricio Ugarte, Daniel Maturana, and Sergio Pizarro.   The SOAR telescope is operated by the Association of Universities for Research in Astronomy, Inc., under a cooperative agreement between the CPNq, Brazil, the National Observatory for Optical Astronomy (NOAO), the University of North Carolina, and Michigan State University, USA.  Finally, we would like to thank an anonymous referee for useful comments and suggestions.


\end{document}